\documentclass[twocolumn,showpacs,preprintnumbers,amsmath,amssymb,floatfix,nofootinbib]{revtex4}

\usepackage[applemac]{inputenc}
\usepackage[T1]{fontenc} 
\pdfoutput=1
\usepackage{amssymb}
\usepackage{bbold}
\usepackage{enumerate}
 \usepackage{amsmath} 
\usepackage{subfigure}
\usepackage{amsfonts} 
\usepackage{bm}
\usepackage[pdftex]{graphicx}
\usepackage{footnote}

\def\beq{\begin{equation}}
\def\eeq{\end{equation}}
\def\bsp{\begin{split}}
\def\esp{\end{split}}
\def\bea{\begin{eqnarray}}
\def\eea{\end{eqnarray}}
\def\ba{\begin{array}}
\def\ea{\end{array}}
\def\nn{\nonumber \\}

\def\eps{\epsilon}

\def\lb{\left(}
\def\rb{\right)}

\def\l.{\left.}
\def\r.{\right.}

\def\pa{\partial}

\def\part{\partial}

\def\Tr{\mbox{Tr}}

\def\ie{{\it i.e. }}

\def\Z{\ensuremath{\mathbb{Z}}}

\def\R{\ensuremath{\mathbb{R}}}

\def\nne{\nonumber\\}
\newcommand\EE{{\mathcal E}}
\newcommand\II{{\mathcal I}}
\newcommand\LL{{\mathcal L}}
\newcommand\MM{{\mathcal M}}
\newcommand\OO{{\mathcal O}}
\newcommand\UU{{\mathcal U}}
\newcommand\VV{{\mathcal V}}
\newcommand\WW{{\mathcal W}}
\newcommand\id{\mathbb{1}} 
\newcommand\bfn{{\mathbf n}}
\newcommand\bfx{{\mathbf x}}
\newcommand\bftau{{\bm{\tau}}}

\usepackage[usenames,dvipsnames]{color}

\usepackage{physics}

\renewcommand{\d}[1]{\ensuremath{\operatorname{d}\!{#1}}}
\begin{document}

\preprint{UdeM-GPP-TH-18-261}
\preprint{arXiv:1805.08038}
\title{Vacuum Decay Induced by False Skyrmions}
\author{\'{E}ric Dupuis$^1$}
\email{eric.dupuis.1@umontreal.ca}

\author{Mareike Haberichter$^{2,3}$}                                                                                                                                                                          
 \email{mareike@math.umass.edu}
 
 \author{Richard MacKenzie$^1$}
\email{richard.mackenzie@umontreal.ca}

\author{M.~B.~Paranjape$^1$}
\email{paranj@lps.umontreal.ca}

\author{U. A. Yajnik$^4$}
\email{yajnik@phys.iitb.ac.in}

\affiliation{$^1$Groupe de Physique des Particules, D\'epartement de physique, Universit\'{e} de Montr\'{e}al, C.~P.~6128, Succursale Centre-ville, Montreal, Qu\'{e}bec, Canada, H3C 3J7}

\affiliation{%
$^2$Department of Mathematics and Statistics, University of Massachusetts, Amherst, Massachusetts 01003-4515, USA
}%
\affiliation{%
$^3$Institut f{\"{u}}r Physik, Universit{\"{a}}t Oldenburg, Postfach 2503 D-26111 Oldenburg, Germany
}%

\affiliation{$^4$Indian Institute of Technology Bombay, Mumbai, India}
\begin{abstract}
We consider the Skyrme model modified by the addition of mass terms which explicitly break chiral symmetry and pick out a specific point on the model's target space as the unique true vacuum.  However, they also allow the possibility of false vacua, local minima of the potential energy.  These false vacuum configurations admit metastable skyrmions, which we call false skyrmions.   False skyrmions can decay due to quantum tunnelling, consequently causing the decay of the false vacuum.  We compute the rate of decay of the false vacuum due to the existence of false skyrmions.   
\end{abstract}


\maketitle


\section{Introduction}
The Skyrme model \cite{Skyrme:1961vq,Skyrme:1962vh} was introduced in the 1960's as a non-linear $\sigma$-model which describe the interaction of low-energy mesons. The model also admits topological solitons (skyrmions) which were interpreted as baryons (for a review see \cite{Gisiger:1998tv}).  However, since that time, skyrmions have found a wide variety of applications in particle physics, cosmology and, more recently, condensed matter physics \cite{Han:2017fyd}.  

The basic topological consideration comes from the classification of mappings of $d$-dimensional configuration space $\R^d$ into the target space of the non-linear $\sigma$-model $\MM$
\beq
\varphi: \R^d\rightarrow \MM
\eeq
subject to the constraint that the field goes to a constant at spatial infinity.   Such a constraint allows us to compactify configuration space from $\R^d$ to $S^d$, the $d$-dimensional sphere, and the homotopy classes are then the homotopy groups
\beq
\{\varphi: S^d\rightarrow \MM\}=\Pi_d(\MM)
\eeq
Of course, the existence of non-trivial homotopy classes does not guarantee the existence of nontrivial finite-energy solutions.  Indeed, in the original work of Skyrme, a four-derivative term (the Skyrme term) was added to prevent the collapse of any topologically nontrivial configuration to a singular configuration \cite{Skyrme:1961vq,Skyrme:1962vh}.  Additionally, the potential on the target manifold must have a global minimum, and for finite energy, the asymptotic constant value to which the Skyrme field  goes, sufficiently quickly, must be the global minimum.

However, one can imagine a Skyrme model where the potential on the target manifold has multiple local minima, \ie one true vacuum as well as one or more false vacua.   Furthermore, one can imagine scenarios where the system is trapped in a metastable false vacuum state \cite{PhysRevLett.37.1378,PhysRevD.15.2922,PhysRevD.14.3568,PhysRevD.23.347,Guth:1982pn}.  Indeed, such a scenario would generically happen in a cosmological phase transition wherein the universe cools down quickly leaving large domains trapped in the false vacuum. Similar scenarios can also occur in condensed matter applications where skyrmions arise.  Then the false vacuum can decay to the true vacuum only from quantum fluctuations and quantum tunnelling transitions.  However, relative to the false vacuum, the Skyrme field admits skyrmion type defects, where the constant value that the Skyrme field goes to corresponds to the value of the field at the false vacuum.   The decay rate of the homogeneous and isotropic, false vacuum configuration was computed by Kobzarev et al and by Coleman and collaborators \cite{Kobzarev:1974cp,Coleman:1978sf,Coleman:1977eu,PhysRevD.16.1762}.  

Here we consider the situation that the false vacuum contains a (false) skyrmion type defect.  The false skyrmion, due to topological exigency, requires the true vacuum point on the target manifold to occur at some place in its interior.  This region of true vacuum would in principle like to grow unboundedly, converting false vacuum to true vacuum.  However, in the models considered in this paper, there is a competition between the energy gained by such a process versus the energy lost by growing the size of the wall.  The volume energy gain scales as $\sim -R^3$ while the energy in the wal increases as $\sim R^2$.  Clearly for large enough $R$ the volume term wins.  However for smaller $R$ the wall term dominates and there can be a potential barrier to the region where the volume term dominates.  We find, in the models that we consider here, that the false skyrmion can only decay through quantum tunnelling transitions through such a barrier, and we compute the corresponding tunnelling rate.  

%
\section{Mass Terms and False skyrmions}
In this paper, we begin with the original notion of skyrmions, topologically nontrivial configurations of a nonlinear $SU(2)$ $\sigma$-model whose field $\UU(\bfx)$ takes values in $\MM = S^3$, the group manifold of $SU(2)$. Finite-energy field configurations must go to a constant ($\UU\to\id$, say) at spatial infinity, so they are maps from compactified physical space ($S^3$) into $\MM$. Since $\Pi_3(S^3)=\Z$, topologically nontrivial field configurations exist and are stabilized, as mentioned above, with the addition of a four-derivative term, giving the Skyrme lagrangian:
\beq\label{L1a}
\LL=  \frac{f_\pi^2}{16} \Tr\left[ \partial_\mu \UU^\dagger\partial^\mu\UU\right]+\frac{1}{32e^2}\Tr\left[\UU^\dagger\partial_\mu\UU,\UU^\dagger\partial_\nu\UU\right]^2.
\eeq
If applied to strong interactions, $f_\pi$ may be interpreted as the pion decay constant; $e$ is a dimensionless constant which can be inferred from scattering data \cite{Adkins:1983ya,Adkins:1983hy,Battye:2005nx}. Throughout this paper, we will use energy and length units $f_\pi/(4e)$ and $2/(ef_\pi)$, respectively, as is common practice \cite{manton_sutcliffe_2004}, with which the Skyrme lagrangian simplifies to
\beq\label{L1ab}
\LL=  \frac{1}{2} \Tr\left[ \partial_\mu \UU^\dagger\partial^\mu\UU\right]+\frac{1}{16}\Tr\left[\UU^\dagger\partial_\mu\UU,\UU^\dagger\partial_\nu\UU\right]^2.
\eeq

The Lagrangian (\eqref{L1a}) exhibits the full $SU(2)\times SU(2)$ chiral symmetry of two-flavour QCD. The chiral $SU(2)\times SU(2)$ with element $(\VV,\WW)$ acts on $\UU$ through the action
\beq
(\VV,\WW):\UU\to \VV^\dagger \UU \WW.
\eeq
The vacuum manifold is the entire target space $\MM$.   In any specific vacuum, chiral symmetry is spontaneously broken to an $SU(2)$ subgroup. For instance, the choice $\UU=\id$ is invariant only under the diagonal subgroup $\VV=\WW$.

We will add a potential to the Lagrangian of the form
\beq
\label{eq-massterm}
\LL_{\text{mass}}=-\frac{1}{4} \left(m_1^2\Tr\left[ \id-\UU\right] +m_2^2\Tr\left[ \id-\UU^2\right]\right)
\eeq
with which $\UU=\id$ is the global minimum-energy configuration, or the true vacuum. More generally, explicit chiral symmetry breaking can be achieved with a mass term of the form\footnote{We could write the potential as a function of $\Tr[\UU]$, but this is completely equivalent, as $\Tr[\UU^n]$ admits a polynomial expansion in terms of $\Tr[\UU]$}
 \beq
 \LL_{\text{mass}}=\sum_k C_k\Tr\left[ \UU^k\right].
 \eeq
Writing
\beq
\label{eq-U}
\UU = e^{i\zeta\hat\bfn\cdot\bftau} = \cos\zeta +i \hat\bfn\cdot\bftau\sin\zeta
\eeq
where $\bftau=(\tau_1,\tau_2,\tau_3)$ are the Pauli matrices, we get
 \beq
  \LL_{\text{mass}}=2\sum_k C_k\cos n\zeta
 \eeq
 which is the cosine Fourier series representation of an arbitrary potential $V(\zeta)$.   In the context of low-energy meson interactions, the only physical constraint on such mass terms is that the pion mass be small.  This condition requires that the curvature  (second derivative) of the potential near its global minimum be small.  In fact, if we  allow mass terms to explicitly break the chiral symmetry completely, then any potential on the three-sphere is permitted.  Given that the pions do not form a perfectly degenerate multiplet, it is clear that even the diagonal symmetry is explicitly broken, although softly.  Hence any kind of soft symmetry breaking terms would in principle be permissible.
 
Allowing such possibilities, it is not unreasonable to imagine a theory with a more elaborate potential, one with several local minima and of course one global minimum.  In this case, one could imagine that through some cooling process the system is trapped in a false, metastable vacuum which would be unstable to quantum tunnelling. The topologically nontrivial nature of the configuration space allows for false skyrmions, ones where the field goes from true vacuum to false as $r$ goes from the origin to infinity. These would decay by tunnelling to a configuration that is classically unstable to infinite expansion and dilution. In fact, it is possible that the false vacuum is long-lived whereas false skyrmions decay rapidly, in which case the very presence of false skyrmions could have a dramatic effect on the overall stability of the system.  We will examine exactly such a possibility with the mass term given in \eqref{eq-massterm}.

\section{Skyrme model with a false vacuum}
The model we consider is the Skyrme model \eqref{L1ab} combined with the potential \eqref{eq-massterm}:
\bea\nonumber
\LL&=&  \frac{1}{2} \Tr\left[ \partial_\mu \UU^\dagger\partial^\mu\UU\right]+\frac{1}{16}\Tr\left[\UU^\dagger\partial_\mu\UU,\UU^\dagger\partial_\nu\UU\right]^2\\
&&\quad -\frac{1}{4} \left(m_1^2\Tr\left[ \id-\UU\right] +m_2^2\Tr\left[ \id-\UU^2\right]\right) .\label{L1}
\eea
Using cosmological language for simplicity, we cam imagine that the universe is trapped in the false vacuum. It will of course eventually decay through quantum tunnelling. As has been observed in other contexts, \cite{Kumar:2010mv,Lee:2013ega,Lee:2013zca}, it is possible that topological objects (skyrmions in the current model) which have true vacuum in their core are formed, and that these objects are classically stable yet unstable due to quantum tunnelling. The key question which we will address is whether the presence of solitons has an important effect on false vacuum instability.

For a static configuration, the energy density corresponding to \eqref{L1} is given by
\bea\label{E}\nonumber
\EE=  \frac{1}{2} \Tr\left[ \partial_i \UU^\dagger\partial_i\UU\right]- \frac{1}{16}
\Tr\left[\UU^\dagger\partial_i\UU,\UU^\dagger\partial_j\UU\right]^2\\
+\frac{1}{4}\left(m_1^2\Tr\left[ \id-\UU\right] +m_2^2\Tr\left[ \id-\UU^2\right]\right) .\nonumber
\eea
Constant field configurations have energy density
\beq
\EE_V= \frac{1}{4}\left(m_1^2\Tr\left[ \id-\UU\right] +m_2^2\Tr\left[ \id-\UU^2\right]\right).
\eeq
Writing $\UU$ as in \eqref{eq-U} gives
\beq
\EE_V\equiv V(\zeta)=m_1^2\sin^2\zeta/2+m_2^2\sin^2\zeta.
\eeq
It is easy to see that $\zeta=0$ is a global minimum with vanishing energy density while if $4m_2^2>m_1^2$ there is a second, local minimum at $\zeta=\pi$ with energy density 
$\EE_V(\pi)=m_1^2$.

\subsection{Metastable false solitons} 
Generically, false solitons in a variety of models can be metastable for a wide range of parameters of the model.  We have analyzed magnetic monopoles, vortices and cosmic strings in the false vacuum, \cite{Kumar:2010mv,Lee:2013ega,Lee:2013zca}.  These situations contain gauge fields, which are absent in the case of skyrmions; however, we find the behaviour is quite similar.  Generally, we have found a simple expression for the energy of the soliton in the so-called thin-wall limit.  In this limit, the soliton profile, which interpolates between the true and false vacua, does so abruptly: it is essentially a bubble of true vacuum embedded in the false vacuum with the transition between the two occurring over a length scale much smaller than the bubble radius $R$.   In this case we find
\beq
E= \alpha R^{d-1} +\frac{\beta}{R^{4-d}} -\epsilon R^{d}
\eeq
where $d$ is the dimension of the space.  The first term corresponds to the energy of the wall which is proportional to its area if $d=3$, its length if $d=2$, {\em etc\/}.  The second term corresponds to the energy in the gauge field. For a monopole in 3 dimensional space it is just the $1/R$ Coulomb energy while for vortices or cosmic strings it is the $1/R^2$ energy  in the magnetic flux tube.  The third term is the energy of the true vacuum within the soliton, which is taken to be negative by normalizing the false vacuum to have zero energy density.  It is clear that such an energy function admits a classically stable soliton solution: the first two terms have a nontrivial minimum.  However, this minimum does not actually guarantee the soliton's existence which must be established by solving the full equations of motion allowing for arbitrary non-spherical variations.  Our numerical analysis makes us quite confident that the corresponding metastable solution does exist.   This minimum is separated by a classical potential barrier from region that is unbounded from below as given by the third term.  

Such thin wall solitons have been shown to exist in the abelian Higgs model in 2+1 dimensions where the corresponding soliton is called a vortex, \cite{Nielsen:1973cs}, in the same model in 3+1 dimensions where the corresponding soliton is called a cosmic string, and in the 't Hooft-Polyakov model  \cite{tHooft:1974kcl,Polyakov:1974ek}, giving rise to magnetic monopoles.  There are also interesting 1-dimensional models where kinks or domain walls have true vacuum inside the wall and false vacuum outside \cite{Dupuis:2015fza,Haberichter:2015xga,Ashcroft:2016tgj}, although these metastable configurations owe there stability to a slightly subtler mechanism.    We will find in this paper that such solutions also exist in models with just scalar fields, such as the Skyrme model \cite{Gisiger:1998tv} although then there is no gauge field energy.  However for the Skyrme model, we will see that the Skyrme term provides the inverse powers of the energy which stabilize the soliton against collapse.  

\section{The skyrmion ansatz \label{sec4}}

\subsection{Energy functional}
We take the field $\UU$ to be that given by the rational map ansatz \cite{Houghton:1997kg,Gisiger:1998tv}
\beq
\UU=e^{if(r)\hat\bfn (\hat \bfx)\cdot\bftau},
\eeq
where $\hat\bfn (\hat \bfx)$ is a mapping of $S^2\to S^2$ that corresponds to the rational map of degree $B$.  The baryon number of the configuration then is also $B$ as long as $f(r)$ interpolates from $\pi$ to 0 for a normal skyrmion, but from $2\pi$ to $\pi$ for a false skyrmion.  The baryon number is an odd function of $f(r)$ but it is invariant under shifts  by $\pi$.  If we want to maintain $f(r)\in [0,\pi]$ then we must replace $f(r)\to \pi-f(r)$ and $\hat\bfn\to -\hat\bfn$.  However, the energy functional, apart from the symmetry-breaking mass terms in the potential is invariant under $f(r)\to f(r)+\pi$, the transformation which exchanges the interior with the exterior. 

The (dimensionless) energy functional is given by
\bea
E=\int d^3r\, \mathcal E &\to&\frac{1}{3\pi}\int\,dr {\bigg(}r^2f^{\prime 2}+2B\left(f^{\prime 2}+1\right)\sin^2f\nne
&&\qquad\qquad+\left.\mathcal{I}\frac{\sin^4f}{r^2}+r^2V(f)\right)\label{Ene}
\eea
where we have rescaled the energy by an additional factor $1/12\pi^2$. 
(This is a fairly common practice \cite{manton_sutcliffe_2004}, motivated by the fact that it simplifies the Skyrme-Faddeev-Bogomolny lower bound on the energy \cite{Skyrme:1962vh,Faddeev:1976pg,Bogomolny:1975de} to $E\ge |B|$. Note that recently stronger lower topological energy bounds have been derived in Refs.~\cite{Harland:2013rxa,Adam:2013tga}.)

\begin{figure}[!htb]
{\includegraphics[totalheight=6.0cm]{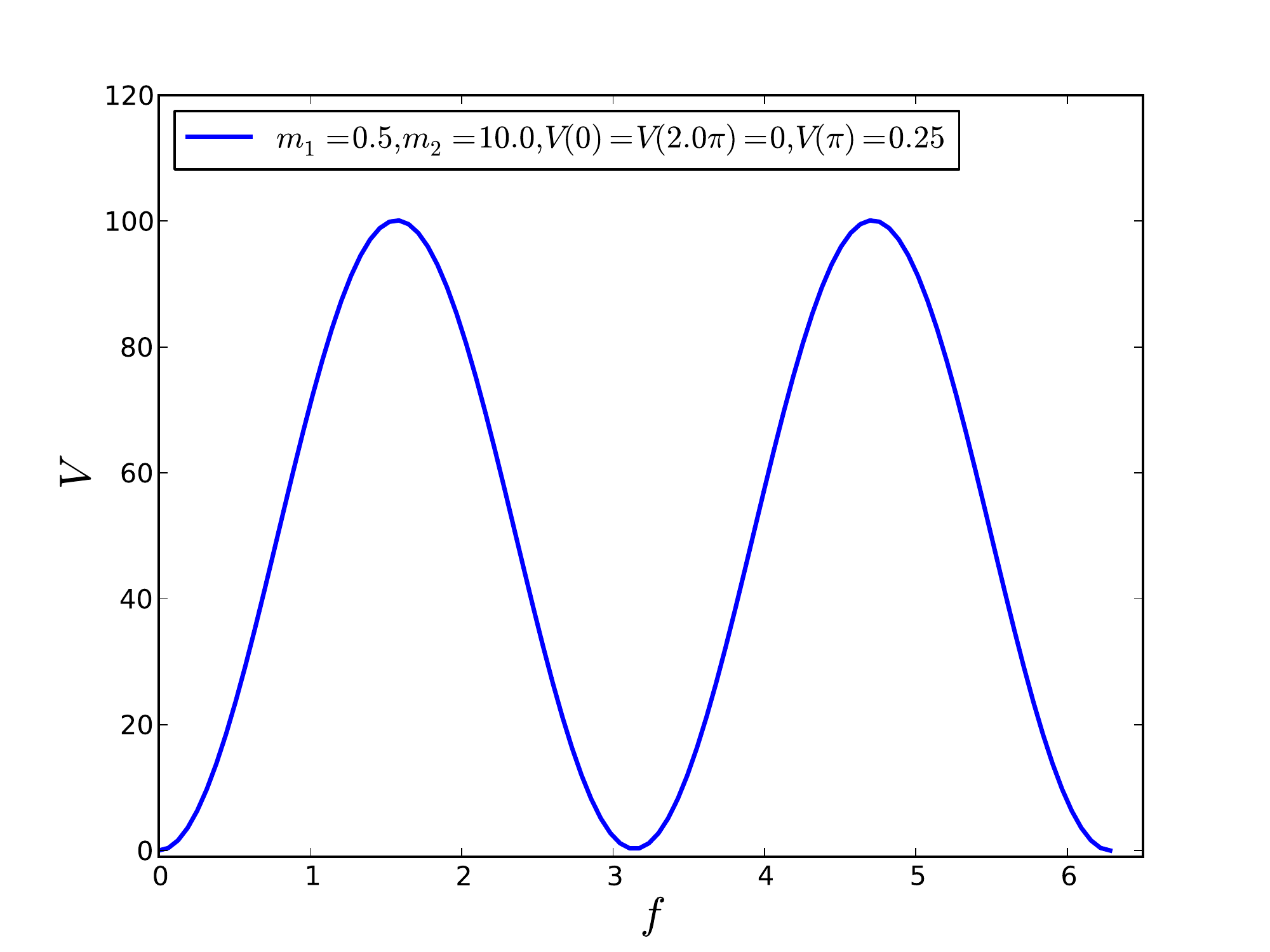}}
\caption{Potential \eqref{Potential} with mass parameters $m_1 =0.5$ and $m_2=10$ as a function of the profile function $f(r)$. $V(\pi)$ is slightly positive, making $f=\pi$ a false vacuum.}
\label{fig1}
\end{figure}

The factor $\mathcal{I}$ appearing in \eqref{Ene} is an integral involving only the rational map and is given in \cite{Houghton:1997kg,Gisiger:1998tv}.  An interesting aspect of $\mathcal{I}$ is that, although evaluated numerically, it is approximately proportional to the baryon number squared in the Skyrme model \cite{Battye:2001qn}: $\mathcal{I}\approx 1.28 B^2$.  This value for $\mathcal{I}$ is actually obtained from numerical studies for $B\in 1\sim 22$, but we will assume that it does not change significantly for much larger baryon number and will therefore extrapolate accordingly.  $\mathcal{I}$ only depends on the rational map and the angular variables and hence cannot depend on the non-standard choice of radial potential \eqref{eq-massterm}.   Explicitly, the potential is
\begin{align}\label{Potential}
V(f)=m_2^2\sin^2f(r)+m_1^2\sin^2\frac{f(r)}{2}.
\end{align}
Throughout the paper we have used $m_1=0.5$, $m_2=10$; the potential for these values is displayed in Fig.~\eqref{fig1}. The potential has a global minimum at $f=0$, where it vanishes,  and a local minimum at $f=\pi$, making the latter a false vacuum. We call the difference in vacuum energy densities $\eps$; for the parameter values used $\eps=0.25$.  (We will in fact give the potential an overall shift so that it is zero at the false vacuum and $-\eps$ in the true vacuum.)

\subsection{Equation of motion}
The equation of motion is given by
\bea\label{eom}
\left(r^2+2B\sin^2f\right)f^{\prime\prime}&+&2f^\prime r +\nne
+\sin2f\left(B\left(f^{\prime2}-1\right)-\frac{\mathcal{I}\sin^2f}{r^2}\right)
&-&\frac{r^2}{2}\frac{\partial V}{\partial f}=0
\eea
with the boundary conditions
\beq\label{BCs}
f(0)=2\pi\,,\quad f(\infty)=\pi .
\eeq
The boundary conditions correspond to a false skyrmion, having the true vacuum at the centre and the false vacuum at infinity.

The equation of motion can be easily solved numerically using MATLAB's \texttt{bvp4c} solver \cite{Gladwell:2003:SOM:862144}. We discretise the two-point boundary problem (\eqref{eom}) on a uniform grid with spatial grid size $r=[0,20]$ and 4000 spatial grid points. We obtain the profile function for the $B=1$ false skyrmion solution by providing the finite difference solver with a crude initial guess satisfying the boundary conditions \eqref{BCs}. False skyrmion solutions for higher baryon numbers $B$, are obtained by simple continuation, that is we increase the baryon number from $B=5$ up to $B=2000$ in steps of $\Delta B=5$ and initialise the solver in each step with the false skyrmion solution obtained in the previous step.

\begin{figure}[!htb]
{\includegraphics[totalheight=7.0cm]{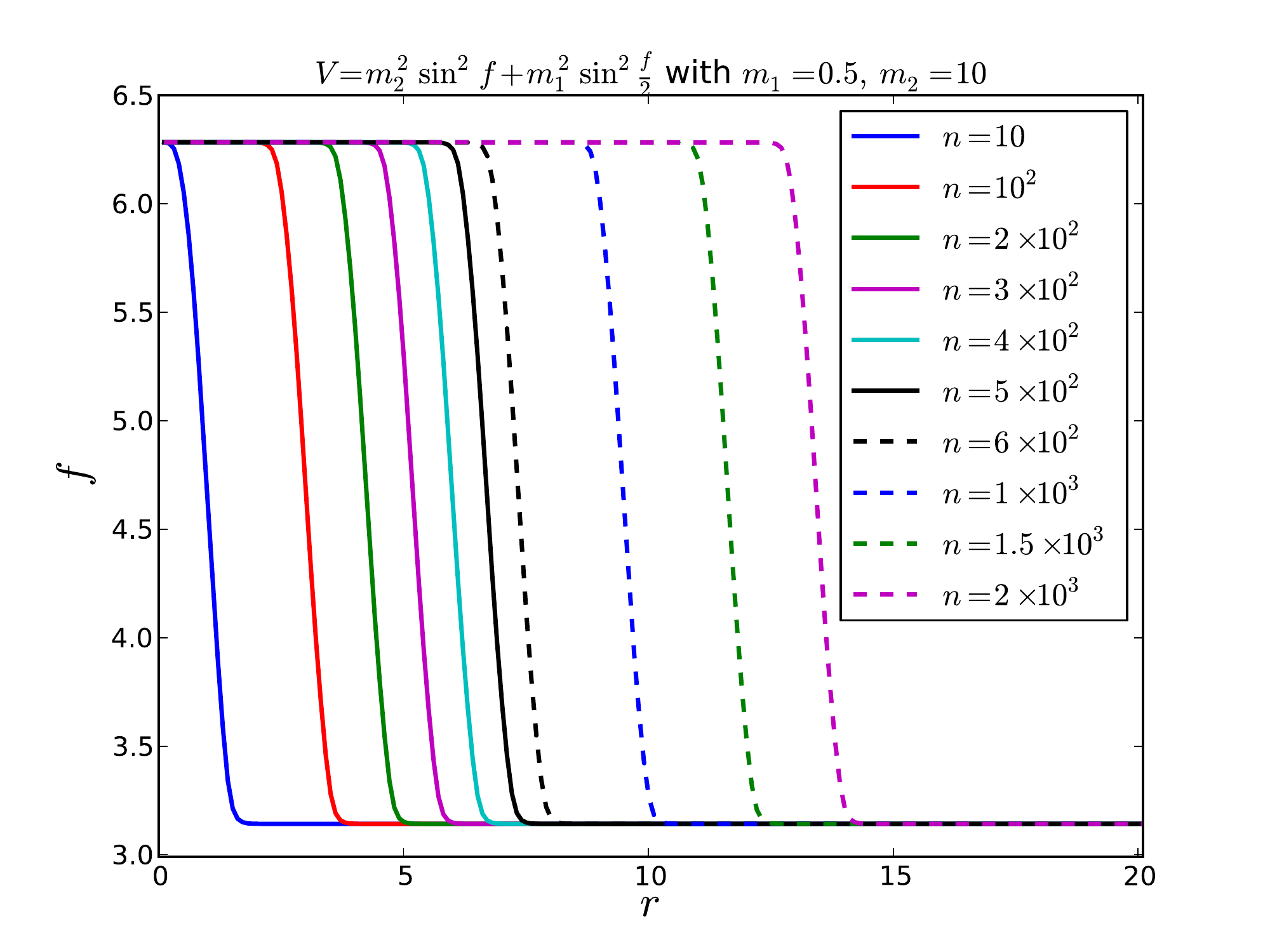}}
\caption{False skyrmion profile functions, $m_2=10$, $m_1=0.5$, $B=n$}\label{fig2}
\end{figure}

\begin{figure}[!h]
{\includegraphics[totalheight=7.0cm]{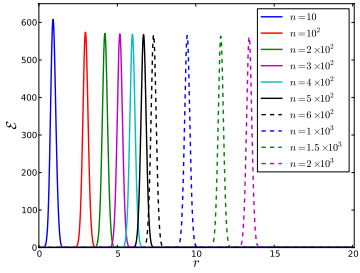}}
\caption{False skyrmion energy density, $m_2=10$, $m_1=0.5$ , $B=n$}\label{fig3}
\end{figure}

\begin{figure}[!h]
{\includegraphics[totalheight=7.0cm]{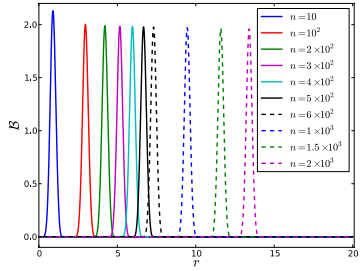}}
\caption{False Skyrmion baryon number density, $m_2=10$, $m_1=0.5$, $B=n$}\label{fig4}
\end{figure}

The profile function $f(r)$ for a wide range of baryon number is displayed in Fig.~\eqref{fig2}, from which it can readily be seen that the profile is indeed of thin-wall type for $B$ sufficiently large.
The energy density and baryon number density are plotted as a function of radius in Fig.~\eqref{fig3} and Fig. \eqref{fig4}, respectively, again displaying the thin-wall nature of the solutions. 

As a measure of the skyrmion's size, we have computed numerically the skyrmion's mean charge radius, which is defined as the square root of the second moment of the baryon number density $\mathcal{B}(r)$, that is,
\beq
\label{eq-dog}
\langle r^2\rangle_Q=\frac{\int d^3r\,r^2\mathcal{B}(r)}{\int d^3r\,\mathcal{B}(r)}\,,
\eeq
with 
\beq
\mathcal{B}(r)=\frac{2n}{\pi}\sin^2f f^\prime\,.
\eeq

In order to allow the interested reader to reproduce our numerical results, we list in Table~\eqref{table-ene_radii} energies and charge radii for false skyrmions for a range of baryon numbers $B$.
\begin{table*}[!t]
\begingroup
\renewcommand{\arraystretch}{3}
 \begin{tabular}{|c|c|c|c|c|c|c|c|c|c|c|c|} \hline
    $B$ & $1$ & $10$ & $100$ &200 &300&400&500&600&1000&1500&2000
     \\ \hline 
    $\displaystyle{E}$ & 2.769 &23.339 & 227.317 &453.752&680.058&906.259&1132.373&1358.405&2261.856&3389.900&4516.776   
    \\ \hline
    $\displaystyle{\langle r^2\rangle^{1/2}_Q}$& 0.370 & 0.984&2.997&4.230&5.179&5.981&6.688&7.327&9.468&11.607&13.415
    \\\hline
  \end{tabular}
\endgroup
\caption{Energies and charge radii for false skyrmion solutions for a range of baryon numbers $B$. Associated profile functions, baryon densities and energy densities can be found in Figs.~\eqref{fig2}-\eqref{fig4}. Recall that energy values are given in units of $12\pi^2$ and we subtracted the energy of the false vacuum.}
\label{table-ene_radii}
\end{table*}

\begin{figure}[!htb]
{\includegraphics[totalheight=7.0cm]{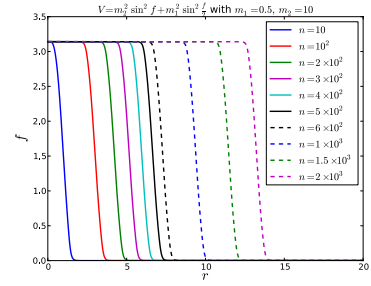}}
\caption{True skyrmion profile functions, $m_2=10$, $m_1=0.5$, $B=n$}\label{figb5}
\end{figure}

\begin{figure}[!h]
{\includegraphics[totalheight=7.0cm]{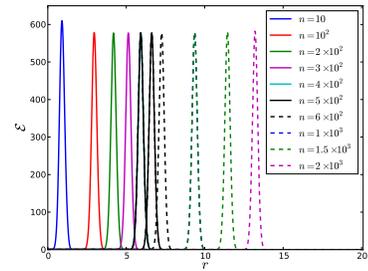}}
\caption{True skyrmion energy density, $m_2=10$, $m_1=0.5$ , $B=n$}\label{figb6}
\end{figure}

The corresponding profiles for true skyrmions are given in Figs.\eqref{figb5} and Fig. \eqref{figb6}.  The profile functions and the energy density are plotted as a function of radius displaying the thin-wall nature of the solutions. 

\begin{figure}[!htb]
{\includegraphics[totalheight=7.0cm]{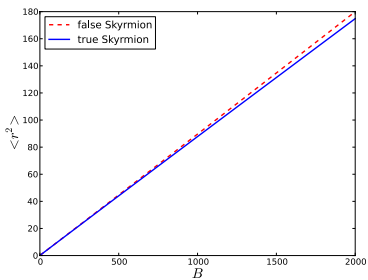}}
\caption{$\langle r^2\rangle$ versus $B$ for the potential with a false skyrmion (blue, solid) and with a true Skyrmion (red, dashed), for the same pion mass. }\label{figb7}
\end{figure}

\begin{figure}[!htb]
{\includegraphics[totalheight=7.0cm]{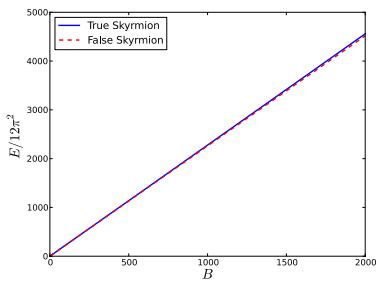}}
\caption{Energy versus $B$ comparing true and false skyrmions.}\label{figb8}
\end{figure}

It is instructive to compare the energy as a function of $B$ for true and false skyrmions in our model.
These are shown in Fig.~\eqref{figb7} and Fig.~\eqref{figb8}.   We notice that they are both thin wall.  This is to be expected, since the potential is essentially a symmetric double well with a relatively small asymmetry; the vacuum energy at the false vacuum is $\approx .25$ while the height of the energy barrier is  $\approx 100$ as can be ascertained from Fig.\eqref{fig1}.  Hence there is not much difference between a true skyrmion and the false skyrmion.  The field $f(r)$ simply eschews the regions where the potential barrier is high, in order to minimize the energy.

\begin{figure}[!h]
{\includegraphics[totalheight=7.0cm]{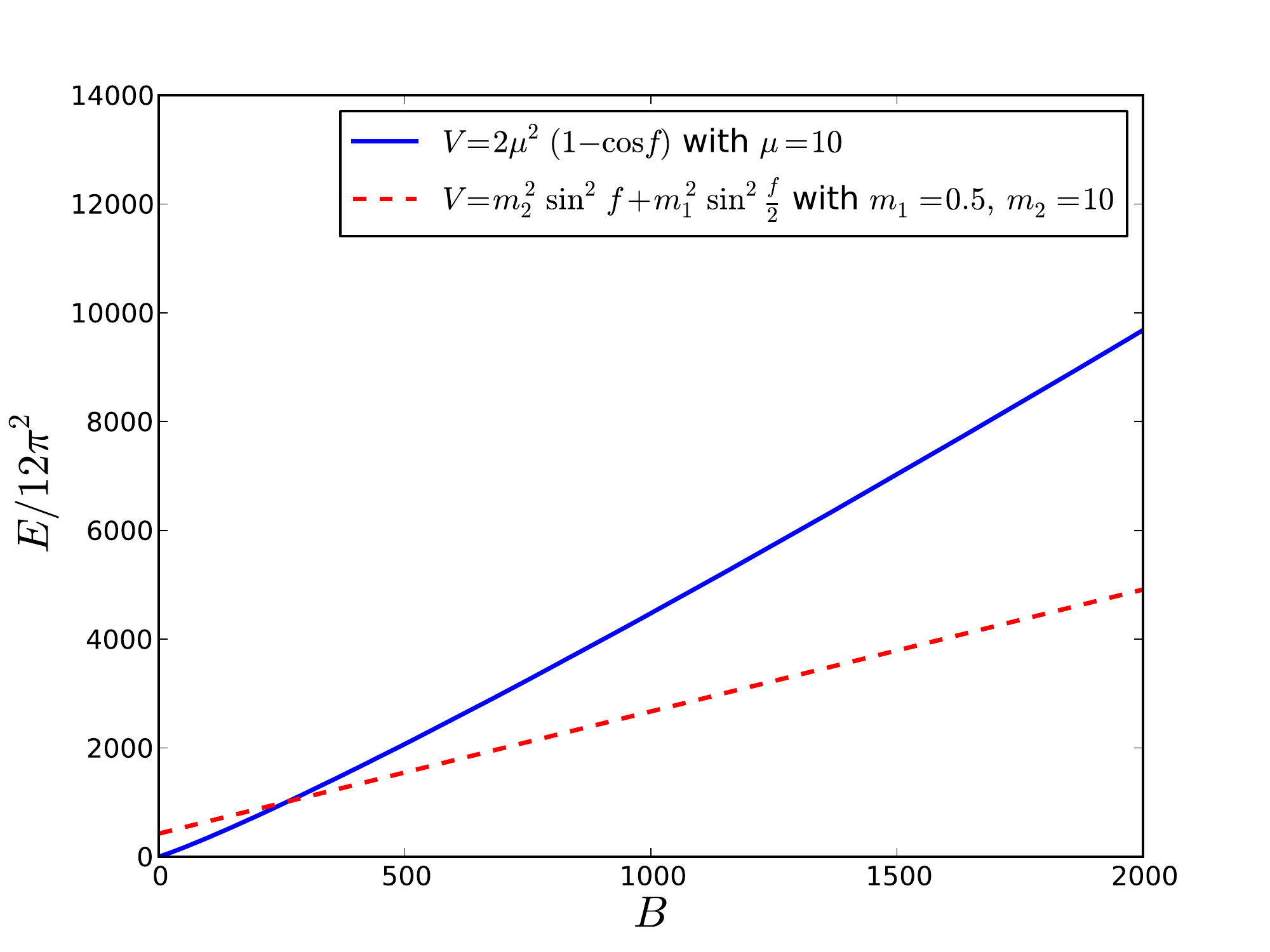}}
\caption{Energy versus $B$ for the potential without a false vacuum (blue, solid) and with a false vacuum (red, dashed), for the same pion mass.}\label{figbb10}
\end{figure}
\begin{figure}[!h]
{\includegraphics[totalheight=7.0cm]{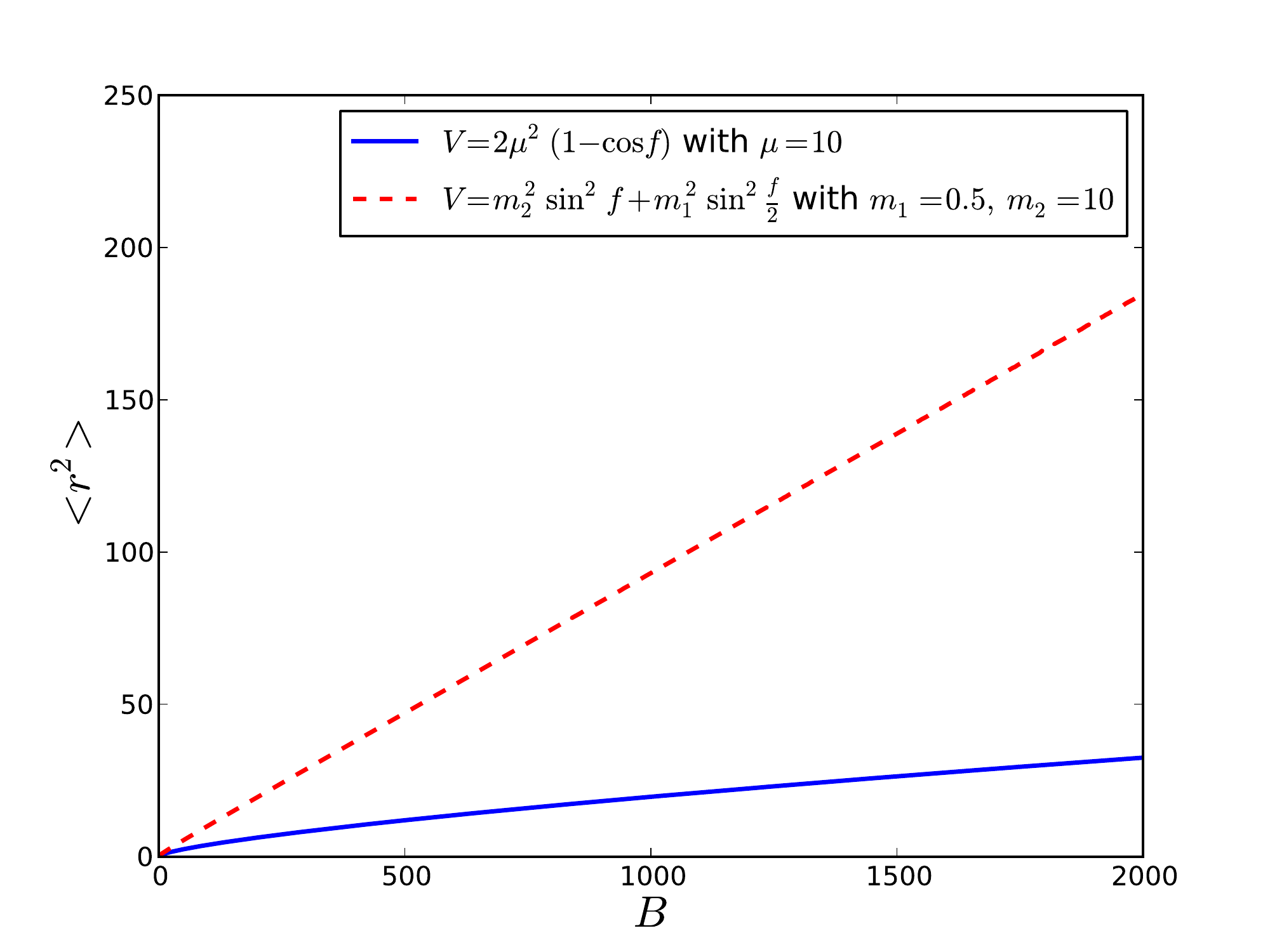}}
\caption{$\langle r^2\rangle$ versus $B$ comparing potentials as in Fig. \eqref{figbb10} }\label{figbb9}
\end{figure}
We can see the effect of the existence of true and false vacua on the skyrmion energy by comparing the model described by \eqref{L1} and the usual model with massive pions, which does not have a false vacuum.  The usual model is obtained by keeping only the first term of \eqref{eq-massterm}.  To make a meaningful comparison, we keep the pion mass (the second derivative of the potential at the vacuum (false or true)) the same.   Fig.\eqref{figbb10} shows a comparison of the energy in the two models.   Fig.~\eqref{figbb9} displays $\langle r^2\rangle$ weighted by the energy density, \eqref{eq-dog} with $\mathcal{B}\to\mathcal{E}$, as a function of the baryon number.
For shell-like skyrmions formed in the model \eqref{L1}, we find a perfect linear correlation, $\langle r^2\rangle \propto B$,  as shown by the dashed line in Fig.~\eqref{figbb9}. The solid line in Fig.\eqref{figbb9} scales more like $B^{2/3}$. Both of these behaviours are as expected, as can be seen by the following argument. In both models,
\bea
\label{eq-fish}
 \langle r^2 \rangle \sim R^2, \quad \text{$R \equiv$  skyrmion radius}.
\eea
However, distinct skyrmion profiles give a different relation between  $B$ and $R$ 
\bea
\label{eq-bird}
 B_{\rm shell} \sim R^2, \qquad B_{\rm ball} \sim R^3.
\eea
Combining \eqref{eq-fish} and \eqref{eq-bird}, we find
\bea	
  \langle r^2 \rangle _{\rm shell} \sim B, \qquad  \langle r^2 \rangle _{\rm ball} \sim B^{2/3}.
\eea
\subsection{Energy}
We have established the existence of thin-wall solutions for skyrmions within the rational map ansatz and evaluated their energy numerically. In fact, the energy can be determined using an analytical approach in the thin wall limit approximatiion. The energy can be written as a sum of contributions from three regions: inside the wall, across the wall and outside the wall. Thus,
\bea
E(R)&=&\int_0^{R-\Delta}dr\mathcal E +\int_{R-\Delta}^{R +\Delta}dr\mathcal E+\int_{R+\Delta}^\infty dr\mathcal E \nne
&=& E_\text{int}+E_{\rm wall}+E_\text{ext}
\label{ER}
\eea
where $R$ is the radius of the thin wall and $\Delta$ is its thickness. Of course, it is assumed that $\Delta\ll R$, an approximation amply justified by the skyrmion profiles shown in Fig.~\eqref{fig2}. In the interior of the skyrmion, $f(r)=2\pi$ which is the true vacuum.  The true vacuum has energy density $-\epsilon$ as the false vacuum is normalized to have zero energy density.  Thus
\beq\label{Eint}
E_\text{int}=-\frac{1}{9\pi}(R-\Delta)^3\epsilon= -\frac{1}{9\pi}R^3\epsilon\lb 1+\OO \lb \frac{\Delta}{R}\rb \rb.
\eeq
In the exterior, the configuration is in the false vacuum, which is normalized to have zero energy density, thus
\beq \label{Eext}
E_\text{ext}=0.
\eeq
For the energy in the wall, $E_{\rm wall}$, we will show that it is easy to obtain an approximate expression for the energy as a function of the radius of the wall.  Consider the equation of motion, multiplied by $f'(r)$
\bea
&~&\left(r^2+2B\sin^2f\right)f^{\prime\prime}f'+2f^{\prime2} r\nonumber\\
&~&+\sin2f\left(B\left(f^{\prime2}-1\right)f'-
\frac{\mathcal{I}\sin^2f}{r^2}\right)f'-\frac{r^2}{2}\frac{\partial V}{\partial f}f'\nonumber\\
&~&=0.
\eea   
Now near the wall, $r$ is large so we can take $r\approx R$.  $f'$ and $f''$ are both of $O(1)$ compared with $R$. Hence the term $2rf'^2$ is negligible compared to $r^2f''f'$, and we will simply drop it.  The rest of the equation can now be integrated, giving
\beq
\frac{R^2 f'^2}{2}+\frac{2B\sin^2f f'^2}{2}-B\sin^2f -\frac{{\mathcal I}\sin^4f}{2R^2} -\frac{R^2V}{2}=0 
\eeq  
where we have normalized the integration constant to vanish.  This allows us to isolate $f'^2$:
\beq\label{fprime2}
f'^2=\frac{2B\sin^2f +({{\mathcal I}\sin^4f}/{R^2}) +R^2V}{{R^2}+2B\sin^2f }
\eeq
which then can be used to obtain the energy in the wall as a quadrature
\bea\label{Ewall}
E_{\rm wall}&=&\frac{2}{3\pi}\int_0^\pi  df \sqrt{\left( 2B\sin^2f +\frac{{\mathcal I}\sin^4f}{R^2} +R^2V \right)}\nonumber\\
&&\qquad\qquad\times\sqrt{({R^2}+2B\sin^2f )},
\eea
Dropping the small symmetry-breaking term in the potential, giving $V\approx m_2^2\sin^2 f$, and making the change of variable $x=\cos f$, we find
\bea
E_{\rm wall}&=&\frac{2}{3\pi}\int_{-1}^1 dx \sqrt{\left( 2B+R^2m_2^2 +\frac{\mathcal I}{R^2}-\frac{\mathcal I}{R^2}x^2\right)}\nne
&&\qquad\qquad\times\sqrt{\left( R^2 +2B -2B x^2\right)}.\label{enqwall}
\eea
This integral can be done analytically and expressed in terms of elliptic functions of the first and second kind; however, the answer is not particularly illuminating.

\begin{figure}[!htb] 
\subfigure[Small-$R$ behaviour; $R_0$ is the false skyrmion radius.\label{fig2-3a}]{\includegraphics[totalheight=4.8cm]{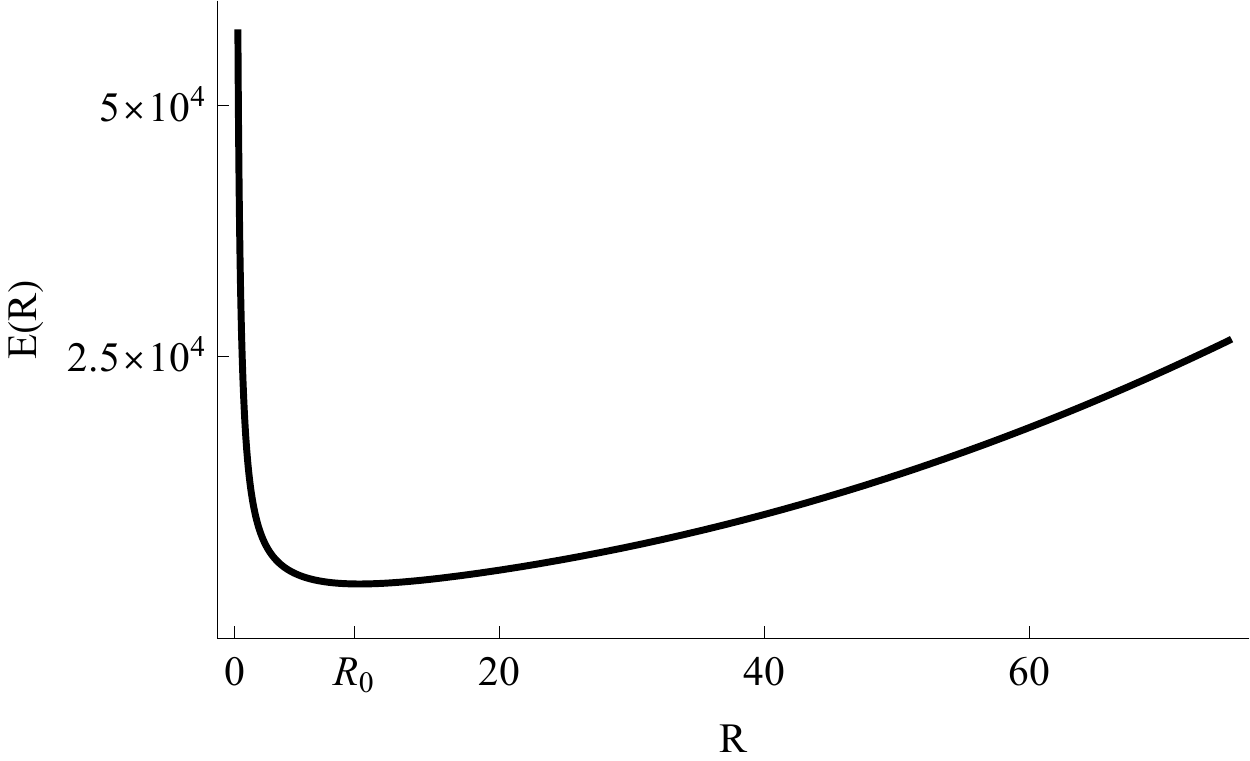}}
\subfigure[Large-$R$ behaviour; $R_1$ is the escape point, for which $E(R_0) = E(R_1)$.\label{fig2-3b}]{\includegraphics[totalheight=4.8cm]{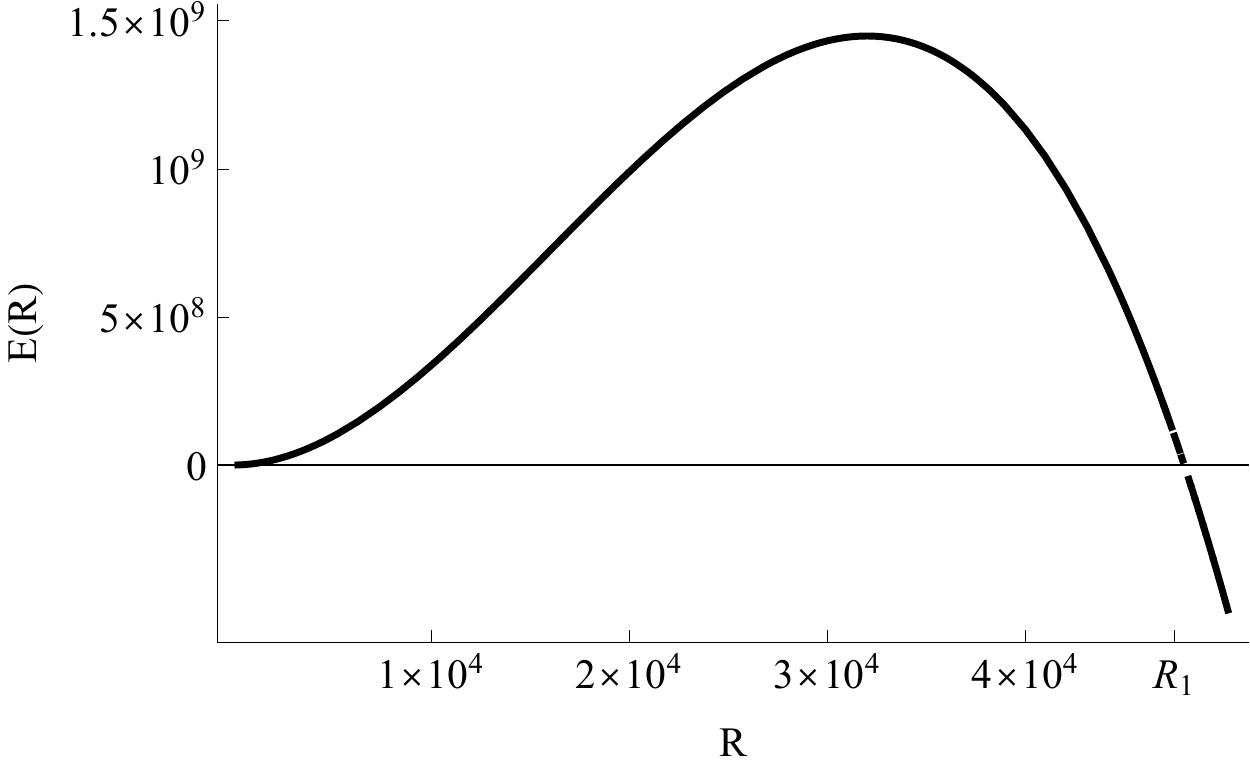}}
\caption{Numerical plot of the potential for thin wall false skyrmions for $N=1000$, $m_1 =0.05$ and $m_2 = 10$.}\label{fig2-3}
\end{figure}

The energy of the configuration as a function of $R$ is given by the sum of \eqref{Eint}, \eqref{Eext} and \eqref{enqwall}; the result is displayed in Fig.~\eqref{fig2-3}.

We notice that $E_{\rm wall}$ contains the crucial terms which behave as $R^2$ and $1/R$ giving the energy of the thin wall skyrmion as
\beq\label{eq-squirrel}
E= \alpha R^{2} +\frac{\beta}{R} -\frac{1}{9 \pi}\epsilon R^3
\eeq
where $\alpha$ gets contributions mostly from both the kinetic term and Skyrme term and $\beta$ gets its contribution largely from the Skyrme term and simple dimensional analysis implies the $1/R$ behaviour, while $\epsilon$ comes from the mass terms and the energy difference between the true and false vacua. Note that this is only an approximate picture to understand why it should make sense to have a metastable skyrmion.  A quick look back to  \eqref{Ewall} shows the energy's dependence on R is not so simple. Nevertheless, the above expression is useful since it encompasses the large $R$ and small $R$ limits.  Clearly the minimum of this potential must exist for $R_0$ such that $R_0\gg \Delta$.

\section{tunnelling decay}

\subsection{False skyrmion decay\label{Sec:false_skyrmion_decay} }
Fig.~\eqref{fig2-3} shows that the thin-wall skyrmion is a meta\-stable solution. The static skyrmion has a radius $R_0$, Fig.~\eqref{fig2-3a}, but there is a larger escape radius $R_1$, Fig.~\eqref{fig2-3b}, and the static energy is equal at both of these radii. Through quantum tunnelling, the metastable skyrmion can expand to radius $R_1$ and become unstable. In what follows, we study the instanton giving this tunnelling effect. 

Considering the field dependence on time, it is simple to show that the only change occurring is a substitution of the radial derivative 
\beq
f'^2 \to \dot{f}^2 + f'^2.
\eeq
We consider the motion of the skyrmion in Euclidean spacetime $(\tau,\bfx)$, restricting our attention to radial fluctuations of the skyrmion. It is clear that such fluctuations yield instantons of lowest actions, and thus dominate the tunnelling decay rate. In fact, we shall treat the radius $R$ of the skyrmion as a collective coordinate, and as the only dynamical degree of freedom. Furthermore, we shall suppose that the profile of the wall remains unchanged in its rest frame as $R(\tau)$ changes. 

We must calculate the action for the time dependent skyrmion.  There are in principle contributions from the three regions, inside, wall and outside, although some of these are trivial or vanish.  The functional form of the energy inside the skyrmion as given by  \eqref{Eint} is not affected by the addition of $\dot{f}$ since the scalar field is constant inside; thus,
\beq
S^E_\text{int} =  \int d \tau E_\text{int}.
\eeq
 As for the energy on the wall, its contribution to the Euclidean action becomes
\bea 
S^E_{\rm wall} \approx \frac{1}{3 \pi} \int d \tau d r \, \left [ R^2 (\dot{f}^2 + f'^2) \lb 1+ 2B \sin^2 f \rb \right. \nne
\left. + \lb 2 B R^2 \sin^2 f + \II\frac{\sin^4 f}{R^2} + R^2 V(f) \rb \right ].
\eea
where we maintain the thin-wall approximation for which $r \approx R$ on the wall.   To compute this term, it is useful to introduce the Gaussian normal coordinate system (see \cite{Dupuis:2017qyo} for an example of the application in the vortex system)
\begin{equation}
ds^2 = d\tau_p^2 - dr_p^2 - {\bar R}^2(\tau_p, r_p) d\theta^2\,,
\end{equation}
where $g_{\tau_p \tau_p} = 1$. In this system, the wall is located at a fixed value  $r_p = \bar{r}_p$ for which ${\bar R}^2(\tau_p, \bar{r}_p) = R^2(\tau_p)$, where $R$ is the physical radius of the wall. The induced metric on the surface of the wall $\Sigma$ is then 
\begin{equation}
ds^2_{(\Sigma)} = d\tau_p^2 - R(\tau_p)^2 d\theta^2. \label{hyperinmetric}
\end{equation}
With these coordinates, the Euclidean action is given by

\bea
S^E_{\rm wall} = \frac{1}{3 \pi} \int d \tau_p d r_p \, \left [ R^2 \lb \frac{\d f}{\d r_p} \rb^2 \lb 1+ 2B \sin^2 f \rb \right . \nne \left . + \lb 2 B R^2 \sin^2 f + \II\frac{\sin^4 f}{R^2} + R^2 V(f) \rb \right].
\label{eq:Swall}
\eea
By the nature of the coordinate system which is co-moving with the wall, the kinetic term is absent: ${d f}/{d t_p} =0$.   Additionally, we assume that ${d f}/{ d r_p}$ is left unchanged compared to its value \eqref{fprime2} when the skyrmion is static. It then follows that 
\bea
S^E_{\rm wall} = \int d \tau_p E_{\rm wall}
\label{eq:Swall2}
\eea
where $E_{\rm wall}$ is given in \eqref{Ewall}. Going back to the ``lab'' frame, we get a relativistic correction which comprises the kinetic term expected with the motion of the wall
\bea
S^E_{\rm wall} = \int d \tau \frac{E_{\rm wall}}{\gamma^E}, \quad \gamma^E = \lb 1 +\dot{R}^2 \rb^{-1/2}.
\eea
The contribution of the region outside the skyrmion still vanishes when the wall becomes dynamic.   The complete Euclidean action is then given by 
\beq
S^E_{\rm sky} =  \int d \tau L_E =  \int_{R_0}^{R_1} \frac{d R}{\dot{R}} \lb \frac{E_{\rm wall}}{\gamma^E} + E_\text{int} \rb.
\label{eq:sky1}
\eeq
We have a single degree of freedom $R$ so we can simply use methods of classical mechanics to reexpress this action. Invariance of the lagrangian under euclidean time translation implies
\bea
 -\pdv{L_E}{\dot{R}} \dot{R} + L_E = \text{constant}.
\eea
Computation of the left-hand side yields
\bea
\gamma^E E_{\rm wall}  + E_\text{int} \equiv E_0.
\label{E0}
\eea
This is of course energy conservation, hence the name $E_0$ for the constant. With this conserved quantity, useful relations to reexpress the action in \eqref{eq:sky1}  can be found. First, $\gamma^E$ is found from \eqref{E0} as
 
%

%
%

\beq
\label{gammaE}
\gamma^E = - \frac{E_\text{int}-E_0}{E_{\rm wall}}.
\eeq
We then can solve for the Euclidean velocity of the wall
\beq
\label{dotR}
 \dot{R} = \sqrt{\lb \frac{E_{\rm wall}}{E_\text{int} -E_0} \rb^2 - 1}.
 \eeq 
In order to compute the tunnelling decay rate, we shall compute the tunnelling exponent which is the difference in action between the instanton configuration and the background skyrmion\footnote{In Coleman's notation, this normalized action would be referred to as $B_{\rm sky} $. Here we use $\tilde{S}_{\rm sky} ^E$ to avoid confusion with the baryon number $B$.} 
 \bea \label{Bsky}
 \tilde{S}^E_{\rm sky} &\equiv& \left. S^E_{\rm sky}\right|_{R(\tau)_{\rm instanton}} - \left. S^E_{\rm sky}\right|_{R_0}\\
				&=&  \int_{R_0}^{R_1} \frac{d R}{\dot{R}} \lb \frac{E_{\rm wall}}{\gamma^E} + E_\text{int} - E_0 \rb.
 \eea
 Using \eqref{gammaE} and \eqref{dotR}, this can be rewritten as
 \beq \label{Bsky2}
 \tilde{S}^E_{\rm sky} = - \int_{R_0}^{R_1} dR \lb E_\text{int} - E_0  \rb \sqrt{\lb \frac{E_{\rm wall}}{E_\text{int} -E_0} \rb^2 - 1}.
 \eeq
 
We will find an analytic expression for $\tilde{S}^E_{\rm sky}$. To do so, we assume $R_1 \gg R_0$. This means the escape radius $R_1$ obtained in the thin-wall approximation must be much larger than the static skyrmion radius $R_0$. The motivation for this approximation is the following.  As a very small difference between false and true vacuum energy density is required from the outset, the volume density contribution rendering the system metastable is very weak, meaning that a large soliton radius is required to before we reach the unstable situation where the energy of the solitons is less than that of the initial, metastable soliton.

 We shall first obtain the desired analytic expression. Afterwards, we will verify that the approximation employed, $R_1\gg R_0 $,  is self-consistent. 
We first write $S^E_{\rm sky}$ using $R_0 \ll R_1$,
\bea
S^E_{\rm sky}   	&=& \int_{R_0}^{R_1} \frac{d R}{\dot{R}} L_E\nne
					 	&\approx& \int_0^{R_1} d R  \left.\lb\frac{L_E}{\dot{R}}\rb\right|_{R\gg R_0} 
\eea
The integrand for $R\gg R_0$ admits a Laurent expansion in powers of $R$ with a finite number of positive powers. The dominant contribution to the integral comes from this region, which can be ascertained by studying the contributions from the interior and from the wall as given by Eqns.(\eqref{Eint},\eqref{enqwall}).  In this region, we can approximate the integrand with its largest polynomial power.  The contribution from the region where  $R\approx R_0$ is finite and negligible compared to the contribution from region of large values of $R$.  Therefore, keeping only largest power of $R$ in the integrand, and integrating from some  large value $R'$ to  $R_1$, where $R'$ satisfies  $R_0\ll R'\ll R_1$, will give a good approximation to the actual integral.  But now, since $R'\ll R_1$, and since furthermore we are integrating only a positive power of $R$, extending the integration all the way down to $R=0$ will give rise to only a tiny contribution to the integral compared to the bulk of the integral coming from $R'\to R_1$.  This approximation to the original integral will simply give an upper bound to the actual integral, with corrections that are small compared to the computed approximate value.  

With this in mind, we wish to simplify $L_E$. There are two contributions, $E_\text{int}$ and $E_{\rm wall}$, which are respectively defined  in \eqref{Eint} and (\eqref{eq:Swall}, \eqref{eq:Swall2}). Keeping only the largest powers of $R$, we obtain
\bea
E_\text{int} - E_0 &\approx& E_\text{int} =  -\frac{1}{3\pi}\frac{\epsilon}{3}  R^3 ,\label{Eintapprox}\\
E_{\rm wall} &\approx&  \frac{R^2}{3 \pi} \int_0^\pi d r_p \lb f'^2 + V(f) \rb \equiv \frac{R^2}{3\pi} \sigma,
\label{Ewallapprox}
\eea
where $\sigma$ is the surface energy density, a quantity which can be computed in our approximation scheme using \eqref{fprime2}:
\bea
\sigma &\equiv& \int_0^\pi d r \lb f'^2 + V(f) \rb  
 {\approx} 2 \int_0^\pi d r \, f'^2 \nn 
 &=& 2 \int_0^\pi d f \, \sqrt{V(f)} = 4 m_2 +\OO\lb \frac{m_1^2}{m_2^2} \rb,
 \label{eq-cat}
\eea
 where again, we have kept only highest powers of $R$ when using \eqref{fprime2}.
Inserting Eqns.(\eqref{Eintapprox}, \eqref{Ewallapprox}) in the action \eqref{Bsky2}, we obtain
 \bea
 \tilde{S}^E_{\rm sky} 	&\approx& - \int_0^{R_1} dR \, E_\text{int} \sqrt{\lb \frac{E_{\rm wall}}{E_\text{int}} \rb^2 - 1}\nne
 				&=& \frac{1}{3\pi} \int_0^{R_1} dR \lb \frac{\epsilon R^3}{3} \rb \sqrt{\lb \frac{\sigma R^2}{(\epsilon/3) R^3} \rb^2 - 1}\nne
 				&=& \frac{1}{3\pi} \frac{\epsilon R_1^4}{3} \int_0^1 d x \, x^3 \sqrt{\frac{1}{x^2} - 1}, \quad x \equiv \frac{R}{R_1}\nne
 				&=& \frac{\epsilon}{144} R_1^4.
 \eea

 We must now evaluate the escape radius $R_1$. We again rely on the approximation $R_0 \ll R_1$ which justifies neglecting small powers of $R$. We search for zeroes of $\dot{R}$ since the boundary conditions of the expanding skyrmion imply, among others, that $\dot{R}|_{R=R_1} = 0$. Going back to the definition \eqref{dotR}, $\dot{R}$ can be approximated with (\eqref{Eintapprox}, \eqref{Ewallapprox})
\bea
\dot{R} = \sqrt{\lb \frac{\sigma R^2}{(\epsilon/3) R^3} - 1 \rb^2}.
\eea
This vanishes for 
\bea
R_1 = 3 \sigma / \epsilon.
\eea
The computation of the action can now be completed 
\bea
\tilde{S}^E_{\rm sky}  = \frac{\epsilon}{144} R_1^3 = \frac{9 \sigma^4}{16 \epsilon^3} \equiv  144 \frac{m_2^4}{m_1^6}.
\label{eq:SEana}
\eea

We now verify our claim that $R_0 \ll R_1$. We have computed $R_1 = 3 \sigma / \epsilon \equiv 12 (m_2 / m_1^2)$ which can be written
\bea
 R_1 = 12 \frac{m_2}{m_1^2} = \frac{12}{m_2} \, \lb \frac{m_2}{m_1}\rb^2 \sim \lb \frac{m_2}{m_1}\rb^2 \gg 1.
 \eea 
for the parameters considered here. Moreover, $R_0$ is the only length scale relevant to describe the false skyrmion and it should obey $R_0 \sim \sqrt{B}$. This behaviour has indeed been demonstrated in Fig.~\eqref{figbb9}. Comparing $R_1$ and $R_0$, we find the relation 
\bea
R_0 = \sqrt{B} \ll \lb \frac{m_2}{m_1} \rb^2 \sim R_1.
\eea
This relation between parameters, $\sqrt{B} \ll  \lb m_2 / m_1 \rb^2$, implying $R_0 \ll R_1$, is observed in solutions presented in Sec.~\eqref{sec4}. Thus, there is a class of metastable skyrmions whose lifetimes are easily computed with the analytical tunneling exponent \eqref{eq:SEana}. The contribution of these defects in determining the false vacuum lifetime is determined by comparing it to other destabilizing effects.

\subsection{False vacuum decay} 
We now turn our attention to false vacuum decay. As mentioned earlier, the false vacuum is given by a constant unitary matrix parameterized as $\UU=e^{i\zeta\hat\bfn\cdot\bftau}$ with $\zeta = \pi$ and then $\hat\bfn$ is irrelevant, $\UU=-\id$.  Deviating from this false vacuum is done only through $\zeta=g(\tau,\bfx)$.  In this given background, a true vacuum bubble can nucleate and induce a phase transition. We assume the associated instanton has the form $\UU= R e^{i g(\tau, \bfx) \hat{n} \cdot\bftau} R^\dagger$  where $R$ denotes a constant, global rotation. Then the Skyrme term is zero and we find a simple scalar field theory with potential $V(g)$
\bea
S^E = \frac{1}{12\pi^2} \int d^4 x  \lb  \pa_\mu g \pa^\mu g + V(g) \rb.
\eea
 In this case, we can use the result of Coleman and collaborators \cite{Coleman:1978sf,Coleman:1977eu,PhysRevD.16.1762}, according to which the nontrivial configuration of minimial action is spherically symmetric. Thus, assuming $g(\rho)$ where $\rho = \sqrt{\tau^2 + |\bfx^2|}$, we obtain the following equation of motion
\beq \label{EOMg}
g'' + \frac{3}{\rho^2} g' - \frac{1}{2}\frac{\pa V(g)}{\pa g} = 0.
\eeq
A thin-wall solution, for which $g$ makes a sharp transition from $\pi$ to $0$, is possible given that $V(g)$ is nearly degenerate, that is, $m_1 / m_2 \ll 1$. We play the same game as for the false skyrmion. Given that $\rho$ is large, we can drop the second term in the equation of motion \eqref{EOMg}. Then, we obtain the first integral
\beq
g'^2 - V(g)  =  - V(\pi).
\eeq
Then the corresponding action is
\bea
\tilde{S}^E_{\rm vac} 	&=&  	\frac{1}{12\pi^2} \int d^4 x  \lb  g'^2 + V(g) - V(\pi) \rb \nne
 				&=& 	\frac{1}{12\pi^2} \int d \Omega \lb  \int_{0}^{\bar \rho - \delta} d \rho \, \rho^3 \lb V(g) - V(\pi) \rb  \right .\nne && \left. \qquad +  \int_{\bar \rho - \delta}^{\bar \rho + \delta} d \rho \, \rho^3 \lb  g'^2 + V(g) - V(\pi) \rb \rb \nne
 				&=& 	\frac{1}{6} \lb   \int_{0}^{\bar \rho - \delta} d \rho \, \rho^3 \lb -m_1^2 \rb   + \int_{\bar \rho - \delta}^{\bar \rho + \delta} d \rho \, \rho^3 \lb 2  g'^2 \rb \rb\nne
 				&\approx& \frac{1}{6} \lb - \frac{\epsilon}{4} \bar \rho^4 + \tilde{\sigma} \bar \rho^3 \rb
\label{Bvac} 				
\eea
where  $\bar\rho$ is the radius of the instanton, the exterior of the instanton gives a vanishing contribution, $\epsilon \equiv m_1^2$ and
\bea
 \tilde{\sigma} &=& \int_{\bar \rho - \delta}^{\bar \rho + \delta} d \rho \, \lb g'^2 + V(g) - V(\pi) \rb  \nne
 &=& 2 \int_0^\pi dg \sqrt{V(g) - V(\pi)} \nne
 &=& 2 m_2 \int_0^\pi d g \sin \lb \frac{g}{2} \rb  +\OO \lb \frac{m_1^2}{m_2} \rb \nne
&=& 4 m_2 +\OO \lb \frac{m_1^2}{m_2} \rb .
\eea
Comparing with \eqref{eq-cat}, we see that $\tilde{\sigma}=\sigma$ to leading order. The action is extremized on physical configurations. Requiring $d \tilde{S}^E_{\rm vac} / d \bar \rho = 0$, the radius of the bounce is obtained
\beq
\tilde \rho = \frac{3 \sigma}{\epsilon} \equiv R_1.
\eeq
This is exactly the escape radius $R_1$ we found for the false skyrmion. The tunnelling exponent is then
\bea
\tilde{S}^E_{\rm vac} = \frac{1}{24} \tilde \rho^4 = \frac{\epsilon}{24} R_1^4  = \frac{9 \sigma^4}{8 \epsilon^3}.
\eea

We note that 
\beq
\tilde{S}^E_{\rm sky} = \frac{\tilde{S}^E_{\rm vac}}{2}.
\eeq
This result is not exclusive to false skyrmions. It was also observed for vortices \cite{Dupuis:2017qyo}. 
This can be understood in the following way.  Once appropriate terms are neglected in the soliton's euclidean action, its disintegration is described by a $O(D-1)$ symmetric vacuum bubble which expands and shrinks in euclidean time, where $D$ is the number of spacetime dimensions. This channel is then compared to conventional vacuum decay given by nucleation of a $O(D)$ symmetric vacuum bubble. Thus, with this universal structure arising for codimension $D-1$ solitons, the general relation $S_{\text{soliton}} = S_{\rm vac}/2 $ is not surprising.  

\subsection{Vacuum decay rates}
For a dilute gas of instantons, the vacuum decay rate (per unit volume) in the semiclassical approximation is given by $\Gamma/V= A e^{-B}\left[1+ \OO(\hbar)\right]$. For the coefficient $A$, the change of variables gives rise to a Jacobian factor which is evaluated in \cite{Coleman:1977eu,PhysRevD.16.1762,paranjape_2017} and yields the decay rate
\begin{equation}
\Gamma =A' L^{(\# \text{zero modes} - 1)}  \left(\frac{\tilde{S}^E}{2\pi}\right)^{(\# \text{zero modes})/2}  e^{-\tilde{S}^E} \,,
\end{equation}
where $A'$ is the determinant excluding the contribution of translational zero modes and hence $V= L^{(\# \text{zero modes} - 1)}$, where $L$ denotes the linear dimension of space.\footnote{This formula only takes into account the translational zero modes, for the (iso-)rotational zero modes, a different factor proportional to the volume of the group will appear, but it is not divergent, and hence does not require special attention.}  We compare the decay rate for skyrmion disintegration and that of regular vacuum decay. The skyrmion  decay rate has to be multiplied by the number $\mathcal{N}$ of skyrmions in the given volume $V$. We suppose a dilute distribution of skyrmions such that inter-skyrmions interactions can be ignored. The ratio of tunnelling rates is then given by
\bea
\frac{\Gamma^{\rm vac}}{\mathcal{N}\Gamma^{\rm sky}} &=& \frac{V A'^{\rm vac}\left(\frac{\tilde{S}^E_{\rm vac}}{2\pi}\right)^{4/2}\, \exp\left(-\tilde{S}^E_{\rm vac}\right)}{\mathcal{N} A'^{\rm sky}\left(\frac{\tilde{S}^E_{\rm sky}}{2\pi}\right)^{1/2}\, \exp\left(-\tilde{S}^E_{\rm sky}\right)} \nne
&=& \frac{ \sqrt{2} A'^{\rm vac} } {(\mathcal{N}/V) A'^{\rm sky} }   \lb \frac{\tilde{S}^E_{\rm vac}}{2 \pi} \rb^{3/2} \exp   \left(-\frac{\tilde{S}^E_{\rm vac}}{2} \right)\nne
\label{eq:ratio}
\eea\\
\noindent
where we have used $\tilde{S}^E_{\rm sky} = \tilde{S}^E_{\rm vac}/2 = 48 m_2^4 / m_1^6 $. $\mathcal{N}/V$ indicates the skyrmion number density. 

The skyrmion number density is assumed to be small enough that there is no significant interaction between the skyrmions.  Thus we assume
\beq
\label{eq-worm}
\mathcal{N}\ll \frac{V}{R_0^3},
\eeq
which simply means that the available volume per skyrmion is much greater than its own volume $\sim R_0^3$.  The size of the skyrmion is fixed by the various parameters in the Lagrangian \eqref{L1} and the total baryon number of the skyrmion.   This can be much smaller than any macroscopic volume $V$ whose decay rate we are interested in, for example the size of the universe.  Hence $V\gg R_0^3$ and ${\cal N}$ can be very large while still satisfying \eqref{eq-worm}.
The number density of topological defects is controlled by the rate of quenching and the correlation length of the fluctuations of the quantum fields as the system passes through the phase transition \cite{Kibble:1976sj,Kibble:1980mv,Zurek:1985qw,Zurek:1993ek}.

We have assumed  from the outset that $\epsilon \ll 1$.  In this limit, we find the tunnelling rate due to false skyrmions to be much larger than the tunnelling rate due to simple, homogeneous vacuum decay, as seen by their ratio, given in \eqref{eq:ratio}, is then very small.  This can occur because $\tilde{S}^E_{\rm vac}$ is very large, particularly as $\epsilon \to 0$, however, as $\tilde{S}^E_{\rm vac}$ becomes large, the tunnelling rate due to both false skyrmion decay and homogeneous vacuum decay both become exponentially small. In turn, this means the ratio, \eqref{eq:ratio} is exponentially small and therefore the phase transition is controlled by the false skyrmion density.  If this is large enough, the vacuum decay will be largely dominated by false skyrmion disintegration. Calculation of the determinant factors $A'^{\rm vac}$ and  $A'^{\rm sky}$ has not been attempted here and is beyond the scope and thrust of this work.

\section{Conclusions}   
Skyrmions are ubiquitous in particle physics and condensed matter physics as solitons which use the topological nature of the space of field configurations and of the space on which they are defined to form topologically stable solitons.  In the case of a false vacuum with the inherent possibility of containing a skyrmion, such a false skyrmion, due to its topological nature, necessarily contains the true vacuum point within its interior, and as such it can induce false vacuum decay.  It is most important to be able to compute the rate at which the induced vacuum decay will occur. 
The present study reveals that false skyrmion decay dominates regular vacuum decay in regions of parameter space where the thin-wall approximation is valid.
\section{Acknowledgments} 
We thank the Ministère des Relations Internationales et de la Francophonie du Gouvernement du Québec under the Cooperation Québec-Maharashtra for continuing financial support. We thank NSERC of Canada generally for financial support and E. D. thanks them specifically for an Alexander Graham Bell Canada Graduate Scholarship.  Some of the work of M.~H. was undertaken as a visiting research scholar at the Department of Mathematics and Statistics, University of Massachusetts, while employed by the University of Oldenburg and financially supported by FP7, Marie Curie Actions, People, International Research Staff Exchange Scheme (IRSES-606096).  We thank the Perimeter Institute for Theoretical Physics, Waterloo, Ontario, the Indian Institute of Technology Bombay, Mumbaï, India, the Inter-University Center for Astronomy and Astrophysics,
Pune, India and the Indian Institute for Science, Education and Research Pune, Pune, India for hospitality during various stages of the
progress of this work. 

\bibliographystyle{apsrev}
\bibliography{Skyrmion-Induced-Vacuum-Decay-ed-v4}

\begin{thebibliography}{39}
\expandafter\ifx\csname natexlab\endcsname\relax\def\natexlab#1{#1}\fi
\expandafter\ifx\csname bibnamefont\endcsname\relax
  \def\bibnamefont#1{#1}\fi
\expandafter\ifx\csname bibfnamefont\endcsname\relax
  \def\bibfnamefont#1{#1}\fi
\expandafter\ifx\csname citenamefont\endcsname\relax
  \def\citenamefont#1{#1}\fi
\expandafter\ifx\csname url\endcsname\relax
  \def\url#1{\texttt{#1}}\fi
\expandafter\ifx\csname urlprefix\endcsname\relax\def\urlprefix{URL }\fi
\providecommand{\bibinfo}[2]{#2}
\providecommand{\eprint}[2][]{\url{#2}}

\bibitem[{\citenamefont{Skyrme}(1961)}]{Skyrme:1961vq}
\bibinfo{author}{\bibfnamefont{T.~H.~R.} \bibnamefont{Skyrme}},
  \bibinfo{journal}{Proc. Roy. Soc. Lond.} \textbf{\bibinfo{volume}{A260}},
  \bibinfo{pages}{127} (\bibinfo{year}{1961}).

\bibitem[{\citenamefont{Skyrme}(1962)}]{Skyrme:1962vh}
\bibinfo{author}{\bibfnamefont{T.~H.~R.} \bibnamefont{Skyrme}},
  \bibinfo{journal}{Nucl. Phys.} \textbf{\bibinfo{volume}{31}},
  \bibinfo{pages}{556} (\bibinfo{year}{1962}), \bibinfo{note}{[,13(1962)]}.

\bibitem[{\citenamefont{Gisiger and Paranjape}(1998)}]{Gisiger:1998tv}
\bibinfo{author}{\bibfnamefont{T.}~\bibnamefont{Gisiger}} \bibnamefont{and}
  \bibinfo{author}{\bibfnamefont{M.~B.} \bibnamefont{Paranjape}},
  \bibinfo{journal}{Phys. Rept.} \textbf{\bibinfo{volume}{306}},
  \bibinfo{pages}{109} (\bibinfo{year}{1998}).

\bibitem[{\citenamefont{Han}(2017)}]{Han:2017fyd}
\bibinfo{author}{\bibfnamefont{J.~H.} \bibnamefont{Han}},
  \bibinfo{journal}{Springer Tracts Mod. Phys.} \textbf{\bibinfo{volume}{278}},
  \bibinfo{pages}{pp.} (\bibinfo{year}{2017}).

\bibitem[{\citenamefont{Frampton}(1976)}]{PhysRevLett.37.1378}
\bibinfo{author}{\bibfnamefont{P.~H.} \bibnamefont{Frampton}},
  \bibinfo{journal}{Phys. Rev. Lett.} \textbf{\bibinfo{volume}{37}},
  \bibinfo{pages}{1378} (\bibinfo{year}{1976}),
  \urlprefix\url{https://link.aps.org/doi/10.1103/PhysRevLett.37.1378}.

\bibitem[{\citenamefont{Frampton}(1977)}]{PhysRevD.15.2922}
\bibinfo{author}{\bibfnamefont{P.~H.} \bibnamefont{Frampton}},
  \bibinfo{journal}{Phys. Rev. D} \textbf{\bibinfo{volume}{15}},
  \bibinfo{pages}{2922} (\bibinfo{year}{1977}),
  \urlprefix\url{https://link.aps.org/doi/10.1103/PhysRevD.15.2922}.

\bibitem[{\citenamefont{Stone}(1976)}]{PhysRevD.14.3568}
\bibinfo{author}{\bibfnamefont{M.}~\bibnamefont{Stone}},
  \bibinfo{journal}{Phys. Rev. D} \textbf{\bibinfo{volume}{14}},
  \bibinfo{pages}{3568} (\bibinfo{year}{1976}),
  \urlprefix\url{https://link.aps.org/doi/10.1103/PhysRevD.14.3568}.

\bibitem[{\citenamefont{Guth}(1981)}]{PhysRevD.23.347}
\bibinfo{author}{\bibfnamefont{A.~H.} \bibnamefont{Guth}},
  \bibinfo{journal}{Phys. Rev. D} \textbf{\bibinfo{volume}{23}},
  \bibinfo{pages}{347} (\bibinfo{year}{1981}),
  \urlprefix\url{https://link.aps.org/doi/10.1103/PhysRevD.23.347}.

\bibitem[{\citenamefont{Guth and Weinberg}(1983)}]{Guth:1982pn}
\bibinfo{author}{\bibfnamefont{A.~H.} \bibnamefont{Guth}} \bibnamefont{and}
  \bibinfo{author}{\bibfnamefont{E.~J.} \bibnamefont{Weinberg}},
  \bibinfo{journal}{Nucl. Phys.} \textbf{\bibinfo{volume}{B212}},
  \bibinfo{pages}{321} (\bibinfo{year}{1983}).

\bibitem[{\citenamefont{Kobzarev et~al.}(1975)\citenamefont{Kobzarev, Okun, and
  Voloshin}}]{Kobzarev:1974cp}
\bibinfo{author}{\bibfnamefont{I.~{\relax Yu}.} \bibnamefont{Kobzarev}},
  \bibinfo{author}{\bibfnamefont{L.~B.} \bibnamefont{Okun}}, \bibnamefont{and}
  \bibinfo{author}{\bibfnamefont{M.~B.} \bibnamefont{Voloshin}},
  \bibinfo{journal}{Sov. J. Nucl. Phys.} \textbf{\bibinfo{volume}{20}},
  \bibinfo{pages}{644} (\bibinfo{year}{1975}), \bibinfo{note}{[Yad.
  Fiz.20,1229(1974)]}.

\bibitem[{\citenamefont{Coleman et~al.}(1978)\citenamefont{Coleman, Glaser, and
  Martin}}]{Coleman:1978sf}
\bibinfo{author}{\bibfnamefont{S.}~\bibnamefont{Coleman}},
  \bibinfo{author}{\bibfnamefont{V.}~\bibnamefont{Glaser}}, \bibnamefont{and}
  \bibinfo{author}{\bibfnamefont{A.}~\bibnamefont{Martin}},
  \bibinfo{journal}{Communications in Mathematical Physics}
  \textbf{\bibinfo{volume}{58}}, \bibinfo{pages}{211} (\bibinfo{year}{1978}).

\bibitem[{\citenamefont{Coleman}(1977)}]{Coleman:1977eu}
\bibinfo{author}{\bibfnamefont{S.}~\bibnamefont{Coleman}},
  \bibinfo{journal}{Physical Review D} \textbf{\bibinfo{volume}{16}},
  \bibinfo{pages}{1248} (\bibinfo{year}{1977}).

\bibitem[{\citenamefont{Callan and Coleman}(1977)}]{PhysRevD.16.1762}
\bibinfo{author}{\bibfnamefont{C.~G.} \bibnamefont{Callan}} \bibnamefont{and}
  \bibinfo{author}{\bibfnamefont{S.}~\bibnamefont{Coleman}},
  \bibinfo{journal}{Phys. Rev. D} \textbf{\bibinfo{volume}{16}},
  \bibinfo{pages}{1762} (\bibinfo{year}{1977}).

\bibitem[{\citenamefont{Adkins et~al.}(1983)\citenamefont{Adkins, Nappi, and
  Witten}}]{Adkins:1983ya}
\bibinfo{author}{\bibfnamefont{G.~S.} \bibnamefont{Adkins}},
  \bibinfo{author}{\bibfnamefont{C.~R.} \bibnamefont{Nappi}}, \bibnamefont{and}
  \bibinfo{author}{\bibfnamefont{E.}~\bibnamefont{Witten}},
  \bibinfo{journal}{Nucl. Phys.} \textbf{\bibinfo{volume}{B228}},
  \bibinfo{pages}{552} (\bibinfo{year}{1983}).

\bibitem[{\citenamefont{Adkins and Nappi}(1984)}]{Adkins:1983hy}
\bibinfo{author}{\bibfnamefont{G.~S.} \bibnamefont{Adkins}} \bibnamefont{and}
  \bibinfo{author}{\bibfnamefont{C.~R.} \bibnamefont{Nappi}},
  \bibinfo{journal}{Nucl. Phys.} \textbf{\bibinfo{volume}{B233}},
  \bibinfo{pages}{109} (\bibinfo{year}{1984}).

\bibitem[{\citenamefont{Battye et~al.}(2005)\citenamefont{Battye, Krusch, and
  Sutcliffe}}]{Battye:2005nx}
\bibinfo{author}{\bibfnamefont{R.~A.} \bibnamefont{Battye}},
  \bibinfo{author}{\bibfnamefont{S.}~\bibnamefont{Krusch}}, \bibnamefont{and}
  \bibinfo{author}{\bibfnamefont{P.~M.} \bibnamefont{Sutcliffe}},
  \bibinfo{journal}{Phys. Lett.} \textbf{\bibinfo{volume}{B626}},
  \bibinfo{pages}{120} (\bibinfo{year}{2005}).

\bibitem[{\citenamefont{Manton and Sutcliffe}(2004)}]{manton_sutcliffe_2004}
\bibinfo{author}{\bibfnamefont{N.}~\bibnamefont{Manton}} \bibnamefont{and}
  \bibinfo{author}{\bibfnamefont{P.}~\bibnamefont{Sutcliffe}},
  \emph{\bibinfo{title}{Topological Solitons}}, Cambridge Monographs on
  Mathematical Physics (\bibinfo{publisher}{Cambridge University Press},
  \bibinfo{year}{2004}).

\bibitem[{\citenamefont{Kumar et~al.}(2010)\citenamefont{Kumar, Paranjape, and
  Yajnik}}]{Kumar:2010mv}
\bibinfo{author}{\bibfnamefont{B.}~\bibnamefont{Kumar}},
  \bibinfo{author}{\bibfnamefont{M.~B.} \bibnamefont{Paranjape}},
  \bibnamefont{and} \bibinfo{author}{\bibfnamefont{U.~A.}
  \bibnamefont{Yajnik}}, \bibinfo{journal}{Phys. Rev.}
  \textbf{\bibinfo{volume}{D82}}, \bibinfo{pages}{025022}
  (\bibinfo{year}{2010}).

\bibitem[{\citenamefont{Lee et~al.}(2013{\natexlab{a}})\citenamefont{Lee, Lee,
  MacKenzie, Paranjape, Yajnik, and Yeom}}]{Lee:2013ega}
\bibinfo{author}{\bibfnamefont{B.-H.} \bibnamefont{Lee}},
  \bibinfo{author}{\bibfnamefont{W.}~\bibnamefont{Lee}},
  \bibinfo{author}{\bibfnamefont{R.}~\bibnamefont{MacKenzie}},
  \bibinfo{author}{\bibfnamefont{M.~B.} \bibnamefont{Paranjape}},
  \bibinfo{author}{\bibfnamefont{U.~A.} \bibnamefont{Yajnik}},
  \bibnamefont{and} \bibinfo{author}{\bibfnamefont{D.-h.} \bibnamefont{Yeom}},
  \bibinfo{journal}{Phys. Rev.} \textbf{\bibinfo{volume}{D88}},
  \bibinfo{pages}{085031} (\bibinfo{year}{2013}{\natexlab{a}}).

\bibitem[{\citenamefont{Lee et~al.}(2013{\natexlab{b}})\citenamefont{Lee, Lee,
  MacKenzie, Paranjape, Yajnik, and Yeom}}]{Lee:2013zca}
\bibinfo{author}{\bibfnamefont{B.-H.} \bibnamefont{Lee}},
  \bibinfo{author}{\bibfnamefont{W.}~\bibnamefont{Lee}},
  \bibinfo{author}{\bibfnamefont{R.}~\bibnamefont{MacKenzie}},
  \bibinfo{author}{\bibfnamefont{M.~B.} \bibnamefont{Paranjape}},
  \bibinfo{author}{\bibfnamefont{U.~A.} \bibnamefont{Yajnik}},
  \bibnamefont{and} \bibinfo{author}{\bibfnamefont{D.-h.} \bibnamefont{Yeom}},
  \bibinfo{journal}{Phys. Rev.} \textbf{\bibinfo{volume}{D88}},
  \bibinfo{pages}{105008} (\bibinfo{year}{2013}{\natexlab{b}}).

\bibitem[{\citenamefont{Nielsen and Olesen}(1973)}]{Nielsen:1973cs}
\bibinfo{author}{\bibfnamefont{H.~B.} \bibnamefont{Nielsen}} \bibnamefont{and}
  \bibinfo{author}{\bibfnamefont{P.}~\bibnamefont{Olesen}},
  \bibinfo{journal}{Nucl. Phys.} \textbf{\bibinfo{volume}{B61}},
  \bibinfo{pages}{45} (\bibinfo{year}{1973}).

\bibitem[{\citenamefont{'t~Hooft}(1974)}]{tHooft:1974kcl}
\bibinfo{author}{\bibfnamefont{G.}~\bibnamefont{'t~Hooft}},
  \bibinfo{journal}{Nucl. Phys.} \textbf{\bibinfo{volume}{B79}},
  \bibinfo{pages}{276} (\bibinfo{year}{1974}).

\bibitem[{\citenamefont{Polyakov}(1974)}]{Polyakov:1974ek}
\bibinfo{author}{\bibfnamefont{A.~M.} \bibnamefont{Polyakov}},
  \bibinfo{journal}{JETP Lett.} \textbf{\bibinfo{volume}{20}},
  \bibinfo{pages}{194} (\bibinfo{year}{1974}), \bibinfo{note}{[Pisma Zh. Eksp.
  Teor. Fiz.20,430(1974)]}.

\bibitem[{\citenamefont{Dupuis et~al.}(2015)\citenamefont{Dupuis, Gobeil,
  MacKenzie, Marleau, Paranjape, and Ung}}]{Dupuis:2015fza}
\bibinfo{author}{\bibfnamefont{E.}~\bibnamefont{Dupuis}},
  \bibinfo{author}{\bibfnamefont{Y.}~\bibnamefont{Gobeil}},
  \bibinfo{author}{\bibfnamefont{R.}~\bibnamefont{MacKenzie}},
  \bibinfo{author}{\bibfnamefont{L.}~\bibnamefont{Marleau}},
  \bibinfo{author}{\bibfnamefont{M.~B.} \bibnamefont{Paranjape}},
  \bibnamefont{and} \bibinfo{author}{\bibfnamefont{Y.}~\bibnamefont{Ung}},
  \bibinfo{journal}{Phys. Rev. D} \textbf{\bibinfo{volume}{92}},
  \bibinfo{pages}{025031} (\bibinfo{year}{2015}).

\bibitem[{\citenamefont{Haberichter et~al.}(2016)\citenamefont{Haberichter,
  MacKenzie, Paranjape, and Ung}}]{Haberichter:2015xga}
\bibinfo{author}{\bibfnamefont{M.}~\bibnamefont{Haberichter}},
  \bibinfo{author}{\bibfnamefont{R.}~\bibnamefont{MacKenzie}},
  \bibinfo{author}{\bibfnamefont{M.~B.} \bibnamefont{Paranjape}},
  \bibnamefont{and} \bibinfo{author}{\bibfnamefont{Y.}~\bibnamefont{Ung}},
  \bibinfo{journal}{J. Math. Phys.} \textbf{\bibinfo{volume}{57}},
  \bibinfo{pages}{042303} (\bibinfo{year}{2016}).

\bibitem[{\citenamefont{Ashcroft et~al.}(2016)\citenamefont{Ashcroft, Eto,
  Haberichter, Nitta, and Paranjape}}]{Ashcroft:2016tgj}
\bibinfo{author}{\bibfnamefont{J.}~\bibnamefont{Ashcroft}},
  \bibinfo{author}{\bibfnamefont{M.}~\bibnamefont{Eto}},
  \bibinfo{author}{\bibfnamefont{M.}~\bibnamefont{Haberichter}},
  \bibinfo{author}{\bibfnamefont{M.}~\bibnamefont{Nitta}}, \bibnamefont{and}
  \bibinfo{author}{\bibfnamefont{M.~B.} \bibnamefont{Paranjape}},
  \bibinfo{journal}{J. Phys.} \textbf{\bibinfo{volume}{A49}},
  \bibinfo{pages}{365203} (\bibinfo{year}{2016}).

\bibitem[{\citenamefont{Houghton et~al.}(1998)\citenamefont{Houghton, Manton,
  and Sutcliffe}}]{Houghton:1997kg}
\bibinfo{author}{\bibfnamefont{C.~J.} \bibnamefont{Houghton}},
  \bibinfo{author}{\bibfnamefont{N.~S.} \bibnamefont{Manton}},
  \bibnamefont{and} \bibinfo{author}{\bibfnamefont{P.~M.}
  \bibnamefont{Sutcliffe}}, \bibinfo{journal}{Nucl. Phys.}
  \textbf{\bibinfo{volume}{B510}}, \bibinfo{pages}{507} (\bibinfo{year}{1998}).

\bibitem[{\citenamefont{Faddeev}(1976)}]{Faddeev:1976pg}
\bibinfo{author}{\bibfnamefont{L.~D.} \bibnamefont{Faddeev}},
  \bibinfo{journal}{Lett. Math. Phys.} \textbf{\bibinfo{volume}{1}},
  \bibinfo{pages}{289} (\bibinfo{year}{1976}).

\bibitem[{\citenamefont{Bogomolny}(1976)}]{Bogomolny:1975de}
\bibinfo{author}{\bibfnamefont{E.~B.} \bibnamefont{Bogomolny}},
  \bibinfo{journal}{Sov. J. Nucl. Phys.} \textbf{\bibinfo{volume}{24}},
  \bibinfo{pages}{449} (\bibinfo{year}{1976}), \bibinfo{note}{[Yad.
  Fiz.24,861(1976)]}.

\bibitem[{\citenamefont{Harland}(2014)}]{Harland:2013rxa}
\bibinfo{author}{\bibfnamefont{D.}~\bibnamefont{Harland}},
  \bibinfo{journal}{Phys. Lett.} \textbf{\bibinfo{volume}{B728}},
  \bibinfo{pages}{518} (\bibinfo{year}{2014}).

\bibitem[{\citenamefont{Adam and Wereszczynski}(2014)}]{Adam:2013tga}
\bibinfo{author}{\bibfnamefont{C.}~\bibnamefont{Adam}} \bibnamefont{and}
  \bibinfo{author}{\bibfnamefont{A.}~\bibnamefont{Wereszczynski}},
  \bibinfo{journal}{Phys. Rev.} \textbf{\bibinfo{volume}{D89}},
  \bibinfo{pages}{065010} (\bibinfo{year}{2014}).

\bibitem[{\citenamefont{Battye and Sutcliffe}(2002)}]{Battye:2001qn}
\bibinfo{author}{\bibfnamefont{R.~A.} \bibnamefont{Battye}} \bibnamefont{and}
  \bibinfo{author}{\bibfnamefont{P.~M.} \bibnamefont{Sutcliffe}},
  \bibinfo{journal}{Rev. Math. Phys.} \textbf{\bibinfo{volume}{14}},
  \bibinfo{pages}{29} (\bibinfo{year}{2002}).

\bibitem[{\citenamefont{Gladwell et~al.}(2003)\citenamefont{Gladwell, Shampine,
  and Thompson}}]{Gladwell:2003:SOM:862144}
\bibinfo{author}{\bibfnamefont{I.}~\bibnamefont{Gladwell}},
  \bibinfo{author}{\bibfnamefont{L.}~\bibnamefont{Shampine}}, \bibnamefont{and}
  \bibinfo{author}{\bibfnamefont{S.}~\bibnamefont{Thompson}},
  \emph{\bibinfo{title}{Solving ODEs with MATLAB}}
  (\bibinfo{publisher}{Cambridge University Press}, \bibinfo{address}{New York,
  NY, USA}, \bibinfo{year}{2003}).

\bibitem[{\citenamefont{Dupuis et~al.}(2017)\citenamefont{Dupuis, Gobeil, Lee,
  Lee, MacKenzie, Paranjape, Yajnik, and Yeom}}]{Dupuis:2017qyo}
\bibinfo{author}{\bibfnamefont{E.}~\bibnamefont{Dupuis}},
  \bibinfo{author}{\bibfnamefont{Y.}~\bibnamefont{Gobeil}},
  \bibinfo{author}{\bibfnamefont{B.-H.} \bibnamefont{Lee}},
  \bibinfo{author}{\bibfnamefont{W.}~\bibnamefont{Lee}},
  \bibinfo{author}{\bibfnamefont{R.}~\bibnamefont{MacKenzie}},
  \bibinfo{author}{\bibfnamefont{M.~B.} \bibnamefont{Paranjape}},
  \bibinfo{author}{\bibfnamefont{U.~A.} \bibnamefont{Yajnik}},
  \bibnamefont{and} \bibinfo{author}{\bibfnamefont{D.-h.} \bibnamefont{Yeom}},
  \bibinfo{journal}{JHEP} \textbf{\bibinfo{volume}{11}}, \bibinfo{pages}{028}
  (\bibinfo{year}{2017}).

\bibitem[{\citenamefont{Paranjape}(2017)}]{paranjape_2017}
\bibinfo{author}{\bibfnamefont{M.~B.} \bibnamefont{Paranjape}},
  \emph{\bibinfo{title}{The Theory and Applications of Instanton
  Calculations}}, Cambridge Monographs on Mathematical Physics
  (\bibinfo{publisher}{Cambridge University Press}, \bibinfo{year}{2017}).

\bibitem[{\citenamefont{Kibble}(1976)}]{Kibble:1976sj}
\bibinfo{author}{\bibfnamefont{T.~W.~B.} \bibnamefont{Kibble}},
  \bibinfo{journal}{J. Phys.} \textbf{\bibinfo{volume}{A9}},
  \bibinfo{pages}{1387} (\bibinfo{year}{1976}).

\bibitem[{\citenamefont{Kibble}(1980)}]{Kibble:1980mv}
\bibinfo{author}{\bibfnamefont{T.~W.~B.} \bibnamefont{Kibble}},
  \bibinfo{journal}{Phys. Rept.} \textbf{\bibinfo{volume}{67}},
  \bibinfo{pages}{183} (\bibinfo{year}{1980}).

\bibitem[{\citenamefont{Zurek}(1985)}]{Zurek:1985qw}
\bibinfo{author}{\bibfnamefont{W.~H.} \bibnamefont{Zurek}},
  \bibinfo{journal}{Nature} \textbf{\bibinfo{volume}{317}},
  \bibinfo{pages}{505} (\bibinfo{year}{1985}).

\bibitem[{\citenamefont{Zurek}(1993)}]{Zurek:1993ek}
\bibinfo{author}{\bibfnamefont{W.~H.} \bibnamefont{Zurek}},
  \bibinfo{journal}{Acta Phys. Polon.} \textbf{\bibinfo{volume}{B24}},
  \bibinfo{pages}{1301} (\bibinfo{year}{1993}).

\end{thebibliography}
\end{document}